\documentclass[aps,prd,onecolumn,groupedaddress,showpacs,nofootinbib,amssymb]{revtex4}
\usepackage[dvips]{graphicx}
\usepackage{amssymb}
\usepackage{amsmath}
\usepackage{graphicx,color}
\usepackage{amsfonts}
\usepackage{bm}
\usepackage{cancel}
\usepackage{comment}

\newcommand\be{\begin{equation}}
\newcommand\ee{\end{equation}}

\allowdisplaybreaks[4]

\begin{document}

\title{Red or Blue Tensor Spectrum from GW170817-compatible Einstein-Gauss-Bonnet Theory: A Detailed Analysis}
\author{V.K. Oikonomou,$^{1,2}$}\,
\email{v.k.oikonomou1979@gmail.com;voikonomou@gapps.auth.gr}
\author{Pyotr Tsyba$^{2}$}\,
\email{pyotrtsyba@gmail.com}
\author{Olga Razina$^{2}$}\,
\email{olvikraz@mail.ru} \affiliation{$^{1)}$ Department of
Physics, Aristotle University of Thessaloniki, Thessaloniki 54124,
Greece\\ $^{2)}$L.N. Gumilyov Eurasian National University -
Astana, 010008, Kazakhstan}

\tolerance=5000

\begin{abstract}
In this work we shall prove that the tensor spectral index of the
primordial tensor perturbations for GW170817-compatible
Einstein-Gauss-Bonnet theories, takes the approximate simplified
form $n_{\mathcal{T}}\simeq 2\left(-1+\frac{1}{\lambda(\phi)}
\right)\epsilon_1$ at leading order, with $\lambda (\phi)$ being a
function of the scalar field which depends on the scalar field
potential and the second derivative of the scalar-Gauss-Bonnet
coupling $\xi''(\phi)$. With our analysis we aim to provide a
definitive criterion for selecting Einstein-Gauss-Bonnet models
that can provide a blue-tilted inflationary phenomenology, by
simply looking at the scalar potential and the scalar-Gauss-Bonnet
coupling. We shall prove this using two distinct approaches and as
we show the tilt of the tensor spectral index is determined by the
values of the potential $V(\phi)$ and of scalar-Gauss-Bonnet
coupling at first horizon crossing. Specifically the blue-tilted
tensor spectral index can occur when $\xi''(\phi_*)V(\phi_*)>0$ at
first horizon crossing.
\end{abstract}


\maketitle

\section{Introduction}

The current interest in theoretical physics is on inflationary
theories
\cite{inflation1,inflation2,inflation3,inflation4,inflation5}
which will be experimentally scrutinized in the next decade.
Indeed, several future experiments will directly seek for hints of
the inflationary theory, either direct like the stage 4 Cosmic
Microwave Background (CMB) experiments
\cite{CMB-S4:2016ple,SimonsObservatory:2019qwx} which will seek
for $B$-modes of inflation, or indirect like the future
gravitational waves experiments
\cite{Hild:2010id,Baker:2019nia,Smith:2019wny,Crowder:2005nr,Smith:2016jqs,Seto:2001qf,Kawamura:2020pcg,Bull:2018lat,LISACosmologyWorkingGroup:2022jok}.
These experiments, apart from constraining, it is also possible to
rule out inflation \cite{Vagnozzi:2022qmc}. The gravitational wave
experiments will be able to discover the existence or not of a
stochastic gravitational waves background with small anisotropies,
which cannot be attributed to astrophysical sources due to the
small anisotropies.

In 2023 an exciting observation from the NANOGrav collaboration
\cite{nanograv} has placed many hopes on the stochastic
gravitational wave background existence. Specifically NANOGrav
\cite{nanograv} confirmed the existence of a stochastic
gravitational wave background and this observation was also
confirmed simultaneously by other pulsar timing arrays (PTA)
experiments \cite{Antoniadis:2023ott,Reardon:2023gzh,Xu:2023wog}.
The absence of large anisotropies \cite{NANOGrav:2023gor}, the
absence of a single supermassive black hole merger event and the
lack of a definitive theoretical solution to the last parsec
problem \cite{Sampson:2015ada} makes the cosmological
interpretation of the signal more likely, at lest for the time
being and many works have appeared in the literature, explaining
the 2023 NANOGrav signal
\cite{sunnynew,Oikonomou:2023qfz,Cai:2023dls,Han:2023olf,Guo:2023hyp,Yang:2023aak,Addazi:2023jvg,Li:2023bxy,Niu:2023bsr,Yang:2023qlf,Datta:2023vbs,Du:2023qvj,Salvio:2023ynn,Yi:2023mbm,You:2023rmn,Wang:2023div,Figueroa:2023zhu,Choudhury:2023kam,HosseiniMansoori:2023mqh,Ge:2023rce,Bian:2023dnv,Kawasaki:2023rfx,Yi:2023tdk,An:2023jxf,Zhang:2023nrs,DiBari:2023upq,Jiang:2023qbm,Bhattacharya:2023ysp,Choudhury:2023hfm,Bringmann:2023opz,Choudhury:2023hvf,Choudhury:2023kdb,Huang:2023chx,Jiang:2023gfe,Zhu:2023lbf,Ben-Dayan:2023lwd,Franciolini:2023pbf,Ellis:2023oxs,Liu:2023ymk,Liu:2023pau,Madge:2023cak,Huang:2023zvs,Fu:2023aab,Maji:2023fhv,Gangopadhyay:2023qjr},
see also
\cite{Schwaller:2015tja,Ratzinger:2020koh,Ashoorioon:2022raz,Choudhury:2023vuj,Choudhury:2023jlt,Choudhury:2023rks,Bian:2022qbh}
for previous works, and also
\cite{Guo:2023hyp,Yang:2023aak,Machado:2018nqk}. In the case that
the stochastic gravitational wave background is due to a
cosmological source, the inflationary perspective of generating
such a signal is significantly constrained because in such a case
a significantly blue-tilted tensor spectral index combined with a
low-reheating temperature is required
\cite{sunnynew,Oikonomou:2023qfz,Vagnozzi:2020gtf,Kuroyanagi:2020sfw,Benetti:2021uea}.

In fact, in some cases it is not necessary to have a strongly
blue-tilted tensor spectral index to explain the NANOGrav signal,
but an abnormal reheating or a post-reheating era is required
\cite{Oikonomou:2023qfz}. The blue-tilted inflationary tensor
spectral index however is a prerequisite to explain the NANOGrav
theories, and there exist theories that can generate a blue-tilted
tensor spectral index, like for example string gas theories
\cite{Kamali:2020drm,Brandenberger:2015kga,Brandenberger:2006pr},
Loop Quantum Cosmology scenarios
\cite{Ashtekar:2011ni,Bojowald:2011iq,Mielczarek:2009vi,Bojowald:2008ik},
non-local gravity models
\cite{Calcagni:2020tvw,Koshelev:2020foq,Koshelev:2017tvv} or
conformal field theories \cite{Baumgart:2021ptt}. One
string-originating theory that can produce a blue-tilted tensor
spectral index is Einstein-Gauss-Bonnet theory, and for mainstream
of articles and reviews see Refs.
\cite{Hwang:2005hb,Nojiri:2006je,Cognola:2006sp,Nojiri:2005vv,Nojiri:2005jg,Satoh:2007gn,Kanti:2015pda,Kanti:1998jd,Kleihaus:2019rbg,Odintsov:2020sqy,Oikonomou:2021kql}
and \cite{reviews1,reviews2,reviews3,reviews4,inflation5}
respectively.

Due to the importance of obtaining a blue-tilted spectral index,
in this article we shall derive and prove in detail that the
tensor spectral index for a general GW170817-compatible
Einstein-Gauss-Bonnet theory can be written in a closed and
simplified form for two distinct approaches, and we shall analyze
the cases when it is blue-tilted or red-tilted and which
parameters or functions determine the sign of the tensor spectral
index at first horizon crossing. As we will show, the size and
magnitude of the second derivative of the Gauss-Bonnet scalar
coupling function $\xi''(\phi)$ at first horizon crossing and the
value of the scalar field potential $V(\phi)$, fully determine the
sign of the tensor spectral index.

\subsection{Inflation with GW170817-compatible Einstein-Gauss-Bonnet Theory and the Tensor Spectral Index}

Before we go to the core of our analysis, let us discuss the
inflationary dynamics of GW170817-compatible Einstein-Gauss-Bonnet
gravity which was developed in
\cite{Odintsov:2020sqy,Oikonomou:2021kql}. The
Einstein-Gauss-Bonnet gravity action is the following,
\begin{equation}
\label{action} \centering
S=\int{d^4x\sqrt{-g}\left(\frac{R}{2\kappa^2}-\frac{\omega}{2}\partial_{\mu}\phi\partial^{\mu}\phi-V(\phi)-\frac{1}{2}\xi(\phi)\mathcal{G}\right)}\,
,
\end{equation}
where $R$ stands for the Ricci scalar, also $\kappa=\frac{1}{M_p}$
where $M_p$ is the reduced Planck mass and furthermore
$\mathcal{G}$ stands for the four dimensional Gauss-Bonnet
invariant which is
$\mathcal{G}=R^2-4R_{\alpha\beta}R^{\alpha\beta}+R_{\alpha\beta\gamma\delta}R^{\alpha\beta\gamma\delta}$
where $R_{\alpha\beta}$ is the Ricci and also
$R_{\alpha\beta\gamma\delta}$ is the Riemann tensor. Let us
consider a flat Friedmann-Robertson-Walker (FRW) geometric
background, with metric,
\begin{equation}
\label{metric} \centering
ds^2=-dt^2+a(t)^2\sum_{i=1}^{3}{(dx^{i})^2}\, ,
\end{equation}
with $a(t)$ denoting as usual the scale factor. By varying the
gravitational action with respect to the metric and with respect
to a time-dependent scalar field, for a FRW metric, the field
equations read,
\begin{equation}
\label{motion1} \centering
\frac{3H^2}{\kappa^2}=\frac{1}{2}\dot\phi^2+V+12 \dot\xi H^3\, ,
\end{equation}
\begin{equation}
\label{motion2} \centering \frac{2\dot
H}{\kappa^2}=-\dot\phi^2+4\ddot\xi H^2+8\dot\xi H\dot H-4\dot\xi
H^3\, ,
\end{equation}
\begin{equation}
\label{motion3} \centering \ddot\phi+3H\dot\phi+V'+12 \xi'H^2(\dot
H+H^2)=0\, .
\end{equation}
For the inflationary era analysis, we shall assume that the
following slow-roll conditions apply,
\begin{equation}\label{slowrollhubble}
\dot{H}\ll H^2,\,\,\ \frac{\dot\phi^2}{2} \ll V,\,\,\,\ddot\phi\ll
3 H\dot\phi\, .
\end{equation}
The speed of the tensor perturbations of the FRW background metric
is \cite{Hwang:2005hb,Odintsov:2020sqy,Oikonomou:2021kql},
\begin{equation}
\label{GW} \centering c_T^2=1-\frac{Q_f}{2Q_t}\, ,
\end{equation}
with $Q_f=8 (\ddot\xi-H\dot\xi)$, $Q_t=F+\frac{Q_b}{2}$,
$F=\frac{1}{\kappa^2}$ and $Q_b=-8 \dot\xi H$. The GW170817 event
imposed the constraint that the gravitational wave speed is nearly
equal to the speed of light, so this means that $Q_f\simeq 0$
which in turn means that $\ddot\xi\simeq H\dot\xi$. By expressing
this in terms of the scalar field, we have,
\begin{equation}
\label{constraint1} \centering
\xi''\dot\phi^2+\xi'\ddot\phi=H\xi'\dot\phi\, ,
\end{equation}
with the ``prime'' indicating differentiation with respect to the
scalar field. Furthermore, if we require that the following
condition holds true,
\begin{equation}\label{firstslowroll}
 \xi'\ddot\phi \ll\xi''\dot\phi^2\, ,
\end{equation}
we can rewrite the constraint (\ref{constraint1}) in the following
way,
\begin{equation}
\label{constraint} \centering
\dot{\phi}\simeq\frac{H\xi'}{\xi''}\, ,
\end{equation}
therefore by combining Eqs. (\ref{motion3}) and (\ref{constraint})
we obtain,
\begin{equation}
\label{motion4} \centering \frac{\xi'}{\xi''}\simeq-\frac{1}{3
H^2}\left(V'+12 \xi'H^4\right)\, .
\end{equation}
Now we further assume that
\begin{equation}\label{mainnewassumption}
\kappa \frac{\xi '}{\xi''}\ll 1\, ,
\end{equation}
\begin{equation}\label{scalarfieldslowrollextra}
12 \dot\xi H^3=12 \frac{\xi'^2H^4}{\xi''}\ll V\, ,
\end{equation}
which will simplify the field equations and we can obtain the
slow-roll indices in closed form. Hence, in view of Eqs.
(\ref{slowrollhubble}), (\ref{constraint}) and
(\ref{scalarfieldslowrollextra}), the field equations become,
\begin{equation}
\label{motion5} \centering H^2\simeq\frac{\kappa^2V}{3}\, ,
\end{equation}
\begin{equation}
\label{motion6} \centering \dot H\simeq-\frac{1}{2}\kappa^2
\dot\phi^2\, ,
\end{equation}
\begin{equation}
\label{motion8} \centering \dot\phi\simeq\frac{H\xi'}{\xi''}\, ,
\end{equation}
and furthermore the condition (\ref{scalarfieldslowrollextra})
becomes,
\begin{equation}\label{mainconstraint2}
 \frac{4\kappa^4\xi'^2V}{3\xi''}\ll 1\, .
\end{equation}
In addition, the differential equation (\ref{motion4}) which
directly relates the functional forms of the  scalar field
potential and the scalar field coupling function of the
Gauss-Bonnet invariant, takes the form,
\begin{equation}
\label{maindiffeqnnew} \centering
\frac{V'}{V^2}+\frac{4\kappa^4}{3}\xi'\simeq 0\, .
\end{equation}
In view of the above, we can obtain the slow-roll indices acquire
a closed and simplified form,
\begin{equation}
\label{index1} \centering \epsilon_1\simeq\frac{\kappa^2\omega
}{2}\left(\frac{\xi'}{\xi''}\right)^2\,
,\,\,\,\epsilon_2\simeq1-\epsilon_1-\frac{\xi'\xi'''}{\xi''^2},\,\,\,
\epsilon_3=0\, ,
\end{equation}
\begin{equation}
\label{index4} \centering
\epsilon_4\simeq\frac{\xi'}{2\xi''}\frac{\mathcal{E}'}{\mathcal{E}}\,
,
\end{equation}
\begin{equation}
\label{index5} \centering
\epsilon_5\simeq-\frac{2\kappa^4\xi'^2V}{3\xi''-4\kappa^4\xi'^2V}\,
,
\end{equation}
\begin{equation}
\label{index6} \centering \epsilon_6\simeq
\epsilon_5(1-\epsilon_1)\, ,
\end{equation}
with $\mathcal{E}=\mathcal{E}(\phi)$ being defined in the
following way,
\begin{equation}\label{functionE}
\mathcal{E}(\phi)=\frac{1}{\kappa^2}\left(
1+72\frac{\epsilon_1^2}{\lambda^2} \right)\, ,
\end{equation}
with the function $\lambda(\phi)$ being defined as follows,
\begin{equation}\label{lambdadef}
\lambda(\phi)=\frac{3\omega}{4\xi''\kappa^2 V}\, .
\end{equation}
Now, in terms of the slow-roll indices, the scalar spectral index
$n_{\mathcal{S}}$, the tensor-to-scalar ratio $r$ and the
tensor-spectral index $n_{\mathcal{T}}$ are,
\begin{equation}
\label{spectralindex} \centering
n_{\mathcal{S}}=1-4\epsilon_1-2\epsilon_2-2\epsilon_4\, ,
\end{equation}
\begin{equation}\label{tensorspectralindex}
n_{\mathcal{T}}=-2\left( \epsilon_1+\epsilon_6 \right)\, ,
\end{equation}
\begin{equation}\label{tensortoscalar}
r=16\left|\left(\frac{\kappa^2Q_e}{4H}-\epsilon_1\right)\frac{2c_A^3}{2+\kappa^2Q_b}\right|\,
,
\end{equation}
where $c_A$ denotes the sound wave speed,
\begin{equation}
\label{sound} \centering c_A^2=1+\frac{Q_aQ_e}{3Q_a^2+
\dot\phi^2(\frac{2}{\kappa^2}+Q_b)}\, ,
\end{equation}
with,
\begin{align}\label{qis}
& Q_a=-4 \dot\xi H^2,\,\,\,Q_b=-8 \dot\xi H,\,\,\,
Q_t=F+\frac{Q_b}{2},\,\,\, Q_c=0,\,\,\,Q_e=-16 \dot{\xi}\dot{H}\,
.
\end{align}
The tensor-to-scalar ratio can be further simplified as follows,
\begin{equation}\label{tensortoscalarratiofinal}
r\simeq 16\epsilon_1\, .
\end{equation}
Regarding the tensor spectral index, we shall demonstrate that it
can be written in the following closed and simple form,
\begin{equation}\label{tensorspectralindexfinal}
n_{\mathcal{T}}\simeq -2\epsilon_1\left ( 1-\frac{1}{\lambda
}+\frac{\epsilon_1}{\lambda}\right)\, ,
\end{equation}
where $\lambda(\phi)$ is defined in Eq. (\ref{lambdadef}). We
shall show that the above form can be obtained by using various
approaches. To start with, let us begin our analysis from the
tensor spectral index of Eq. (\ref{tensorspectralindex}) and focus
on the slow-roll index $\epsilon_6$ defined in Eq. (\ref{index6}).
So we start from,
\begin{equation}\label{epsilon6index1}
\epsilon_6=-\frac{2\kappa^4\xi'^2\left(1-\frac{1}{2}\kappa^2\omega
\left(\frac{\xi'}{\xi''} \right)^2
\right)}{\xi''^2\left(\frac{3}{\xi''V}-4\kappa^4\left(\frac{\xi'}{\xi''}\right)^2
\right)}\, ,
\end{equation}
which can be written in terms of the slow-roll index $\epsilon_1$
as follows,
\begin{equation}\label{epsilon6index2}
\epsilon_6=-\frac{\frac{4\kappa^2}{\omega}\epsilon_1\left(1-\epsilon_1
\right)}{\frac{3}{\xi''V}-\frac{8\kappa^2\epsilon_1}{\omega}}\, ,
\end{equation}
or equivalently,
\begin{equation}\label{epsilon6index2}
\epsilon_6=-\frac{\epsilon_1(1-\epsilon_1)}{\frac{3\omega}{4\kappa^2\xi''V}-2\epsilon_1}
\, .
\end{equation}
Thus by inserting the above in Eq. (\ref{tensorspectralindex}) we
get,
\begin{equation}\label{tensorspectralindexfinal1}
n_{\mathcal{T}}\simeq -2\left
(\epsilon_1-\frac{\epsilon_1(1-\epsilon_1)}{\frac{3\omega}{4\kappa^2\xi''}-2\epsilon_1}
\right)\, ,
\end{equation}
and since $\epsilon_1\ll 1$ we expand the above for small values
of $\epsilon_1$ and we get,
\begin{equation}\label{tensorspectralindexfinal2}
n_{\mathcal{T}}\simeq 2\left(-1+\frac{1}{\lambda}
\right)\epsilon_1-\frac{2(\lambda-2)}{\lambda^2}\epsilon_1^2+...-2^{n-1}\frac{\lambda-2}{\lambda^n}\epsilon_1^n\,
.
\end{equation}
Thus at first order, the tensor spectral index has the following
form,
\begin{equation}\label{firstordertensorspectralindex}
n_{\mathcal{T}}\simeq 2\left(-1+\frac{1}{\lambda}
\right)\epsilon_1\, .
\end{equation}
We can arrive at the same expression by using a different
approach, starting from Eq. (\ref{tensorspectralindex}) and using
Eq. (\ref{index6}), and then we get approximately,
\begin{equation}\label{oneneweqn}
n_{\mathcal{T}}\simeq
-2\left(\epsilon_1+\epsilon_5-\epsilon_5\epsilon_1 \right)\, ,
\end{equation}
and it can be proven that $\epsilon_5$ takes a simpler form than
the one in Eq. (\ref{index5}), since,
\begin{equation}\label{rewrittemepsilon5}
\epsilon_5=-\frac{\epsilon_1}{\frac{3\omega}{\xi''V4\kappa^2}-2\epsilon_1}\,
.
\end{equation}
Due to the approximation we assumed earlier, namely
$\frac{4\xi'^2\kappa^4V}{3\xi''}\ll 1$, we get,
\begin{equation}\label{approxepsilon5}
\epsilon_5\simeq
-\frac{2\kappa^4\xi'^2V}{3\xi''}=-\frac{\epsilon_1}{\lambda}\, ,
\end{equation}
thus, the tensor spectral index (\ref{oneneweqn}) acquires the
following form,
\begin{equation}\label{tensoapproach2final}
n_{\mathcal{T}}\simeq
-2\epsilon_1(1-\frac{1}{\lambda})-\frac{2\epsilon_1^2}{\lambda}\,
,
\end{equation}
which is identical to the expression given in Eq.
(\ref{firstordertensorspectralindex}) if terms of the order $\sim
\mathcal{O}(\epsilon_1)$ are kept in the expansion. Thus we proven
using two distinct approaches that the tensor spectral index at
first order approximation can take the simplified form given in
Eq. (\ref{firstordertensorspectralindex}). It is apparent that the
tilt of the spectral index is critically affected by the values of
the function $\lambda(\phi)$ at first horizon crossing, since in
our formalism $\epsilon_1$ is always positive. Thus the following
cases determine the tilt of the tensor spectral index,
\begin{equation}\label{interactionterm2}
n_{\mathcal{T}}\simeq\left\{
\begin{array}{c}
  -2\epsilon_1,\,\,\,\mathrm{when}\,\,\,|\lambda(\phi_*)|\gg 1  \\
 \frac{2\epsilon_1}{\lambda},\,\,\,\mathrm{when}\,\,\,|\lambda(\phi_*)|\ll
 1\\
  0,\,\,\,\mathrm{when}\,\,\,|\lambda(\phi_*)|\simeq 1 \, ,\\
  \frac{\epsilon_1(1-\lambda)}{\lambda}>0,\,\,\,\mathrm{when}\,\,\,\lambda(\phi_*)<1 \, ,
\end{array}\right.
\end{equation}
where $\phi_*$ is the value of the scalar field at the first
horizon crossing. Thus the strongly blue-tilted tensor spectral
index can occur when $|\lambda(\phi_*)|\ll 1$ and when $\lambda
(\phi_*)>0$, and also in the case that $\lambda<1$ the tensor
spectral index is also blue-tilted but not necessarily strongly
blue-tilted. Now it is apparent that the values and the sign of
the function $\lambda (\phi)$ are determined by the values and
signs of $\xi''(\phi)$ and of $V(\phi)$ at the first horizon
crossing. Thus the blue-tilted tensor spectral index can occur
when $\xi''(\phi_*)V(\phi_*)>0$ at first horizon crossing. Clearly
this is more or less model dependent, but our analysis provides
insights in the analysis of inflationary phenomenology of
Einstein-Gauss-Bonnet models, because models with positive
potentials must have $\xi''(\phi)>0$ in order to generate a
blue-tilted tensor spectral index and also models with negative
values of the potential at first horizon crossing must have
$\xi''(\phi_*)<0$ in order to generate a blue-tilted tensor
spectral index.

Let us consider in brief some illustrative examples, the
phenomenology of which is well studied in the literature and these
provide a viable inflationary phenomenology
\cite{Oikonomou:2021kql}. Consider the following functional form
of the scalar coupling function $\xi(\phi)$,
\begin{equation}
\label{modelApower} \xi(\phi)=\beta  \left(\frac{\phi
}{M}\right)^{\nu }\, ,
\end{equation}
with $\beta$ being a dimensionless parameter, and $M$ is a free
parameter with mass dimensions $[m]^1$. By combining Eqs.
(\ref{modelApower}) and (\ref{maindiffeqnnew}) and solving the
differential equation (\ref{maindiffeqnnew}), we get the following
scalar field potential,
\begin{equation}
\label{potApower} \centering V(\phi)=\frac{3}{3 \gamma  \kappa
^4+4 \beta  \kappa ^4 \left(\frac{\phi }{M}\right)^{\nu }} \, ,
\end{equation}
with $\gamma$ being a dimensionless integration constant. Since in
this case, we always have $\xi''(\phi)>0$, the sign of the tensor
spectral index will be determined by the scalar potential value,
which is positive in order for a viable phenomenology to be
ensured. Indeed a viable phenomenology is guaranteed for the
following values of the free parameters \cite{Oikonomou:2021kql}
$\mu=2.09\times 10^{-22}\times \kappa$,
$\beta=[0.000025,1.69055]$, $\gamma=10^{200}$, for $N=[50,63]$
$e$-foldings, for which the corresponding tensor spectral index
values are in the range $n_{\mathcal{T}}=[0.006,1.108]$. Thus for
these values the scalar potential is positive, and as we expected,
the tensor spectral index is blue-tilted. Another model of this
sort is obtained for the scalar coupling function $\xi(\phi)$
having the form $\xi(\phi)=\beta  \exp \left(\left(\frac{\phi
}{M}\right)^2\right)$, where $\beta$ is a dimensionless parameter,
and $M$ is a free parameter with mass dimensions $[m]^1$. By
combining $\xi (\phi)$ and Eq. (\ref{maindiffeqnnew}) and by
solving the differential equation (\ref{maindiffeqnnew}) we obtain
the following scalar potential,
\begin{equation}
\label{potA} \centering V(\phi)=\frac{3}{3 \gamma  \kappa ^4+4
\beta  \kappa ^4 e^{\frac{\phi ^2}{M^2}}} \, ,
\end{equation}
where $\gamma$ is a dimensionless integration constant. In this
case we have $\xi''(\phi)>0$ if $\beta>0$ and $\xi''(\phi)<0$ if
$\beta<0$, and the same applies for the scalar potential, which is
however affected by the free parameter $\gamma$ too. Thus in this
case, the product $\xi(\phi_*)V(\phi_*)$ at the first horizon
crossing will determine the sign of the tensor spectral index.
Therefore, one must find which values of the free parameters
guarantee a viable inflationary phenomenology, and the condition
$\xi(\phi_*)V(\phi_*)>0$ can easily be examined. In this case, a
viable phenomenology is guaranteed for the following values of the
free parameters \cite{Oikonomou:2021kql} $\mu=[22.091,22.0914]$,
$\beta=-1.5$, $\gamma=2$, for $N=60$ $e$-foldings. Thus both the
potential and the second derivative of the scalar Gauss-Bonnet
coupling function are negative, thus the tensor spectral index is
positive. Indeed, as it was shown in \cite{Oikonomou:2021kql} for
these values of the free parameters we have
$n_{\mathcal{T}}=[0.378,0.379]$, hence it is positive and also one
may verify that $\xi(\phi_*)V(\phi_*)>0$. Therefore, with our
analysis one may know beforehand which Einstein-Gauss-Bonnet model
may yield a blue-tilted tensor spectral index. However, we need to
note that while the examples we presented can predict a blue
tensor tilt, it is not clear whether those, or other model in the
class, can provide the high values of the tensor spectral index
required to explain PTA data \cite{sunnynew,NANOGrav:2023hvm}.
This feature can be model dependent, so it is left as a question
to the reader and for future work.

\section{Conclusions}

In this paper we focused on the issue when the tensor spectral
index of the primordial tensor perturbations for
GW170817-compatible Einstein-Gauss-Bonnet theories is blue-tilted
or red-tilted. With our analysis we aimed to provide a definitive
 and simple criterion for appropriately selecting Einstein-Gauss-Bonnet models that can
provide a blue-tilted inflationary phenomenology, by simply
selecting appropriately the scalar potential and the
scalar-Gauss-Bonnet coupling. The importance of a blue-tilted
inflationary phenomenology is great, after the 2023 NANOGrav
observation of the stochastic gravitational wave background. Using
the GW170817-compatible formalism of Einstein-Gauss-Bonnet
inflationary theories, we showed, by using two distinct
approaches, that at leading order, the tensor spectral index takes
the approximate simplified form $n_{\mathcal{T}}\simeq
2\left(-1+\frac{1}{\lambda(\phi)} \right)\epsilon_1$ at leading
order, with $\lambda (\phi)$ being a function of the scalar field
which depends on the scalar field potential and the second
derivative of the scalar-Gauss-Bonnet coupling $\xi''(\phi)$.
Since in our formalism, the slow-roll index $\epsilon_1$ is always
positive, the tilt of the spectral index depends on the sign of
the function $\lambda (\phi)$ at the first horizon crossing, which
in turn depends on the values of the potential $V(\phi)$ and of
scalar-Gauss-Bonnet coupling at first horizon crossing. As we
showed, when $\lambda (\phi_*)\ll 1$ and when $\lambda (\phi_*)<
1$, the tensor spectral index is blue-tilted, while when
$\lambda(\phi_*)\simeq 1$, and $\lambda(\phi_*)> 1$ the tilt is
red, where $\phi_*$ the value of the scalar field at the first
horizon crossing. In terms of the scalar field potential and the
scalar-Gauss-Bonnet coupling, the blue-tilted tensor spectral
index can occur when $\xi''(\phi_*)V(\phi_*)>0$ at first horizon
crossing. Hence our analysis provides insights in the study of
Einstein-Gauss-Bonnet inflationary theories, making it possible to
chose beforehand blue-tilted inflationary phenomenologies.

\section*{Acknowledgments}

This research has been is funded by the Committee of Science of
the Ministry of Education and Science of the Republic of
Kazakhstan (Grant No. AP14869238).

\end{document}